# Completing the dark matter solutions in degenerate Kaluza-Klein theory

Trevor P. Searight[a]
*London, United Kingdom*

A complete set of wave solutions is given for the weak field in a Kaluza-Klein theory with degenerate metric. In the five-dimensional version of this theory electromagnetism is described by two vector fields, and there is a reflection symmetry between them which unifies them with gravitation; wave behaviour in the extra dimension has been interpreted as dark matter. Here three independent dark matter solutions are found, and for two of them it is shown how they must be combined into a single solution in order to obey the reflection symmetry. The unification is also expanded to six dimensions to prepare the way to include further forces.

## I. INTRODUCTION

History teaches that one way to choose between the myriad of possible physical theories is to look for a suitable invariance (or symmetry) – gauge invariance and coordinate invariance being obvious examples. Furthermore, the stronger the invariance, the stronger the theory – the progression from special relativity to general relativity being a case in point. Despite the ongoing challenges surrounding general relativity, such as the existence of singularities, few would question the success of the theory, or doubt that it is part of nature.

In Ref. 1 this author showed that there is a possible unification of gravitation and electromagnetism along the lines of Kaluza-Klein theory but using a degenerate metric in five dimensions. This theory differs from standard Kaluza-Klein theory in that electromagnetism is not combined with gravitation in a single metric, but instead is described by two vector fields with a reflection symmetry between them, such that charge may either be the fifth component of 5-velocity or the fifth component of 5-momentum. One vector field interacts with matter, and is termed "radiant light", while the other does not, and is termed "dark light"; there is also a scalar field which is equivalent to the Brans-Dicke scalar field[2]. In the degenerate theory fields are allowed to vary in the fifth dimension – albeit in a limited way that is determined by the field equations – and in particular they can have wave behaviour in the fifth dimension. This wave behaviour has been interpreted as dark matter: the oscillation gives the field positive energy and so acts as a source of gravitation, but there is no direct interaction with ordinary matter due to an averaging over the wave. In Ref. 1 a single dark matter solution was found by solving the two vector field equations and assuming a flat gravitational field – an assumption that was made for reasons of simplicity. Now a more systematic approach is used to solve for dark matter by looking at the whole field.

This paper is divided into seven sections. Section II sets out the basics of degenerate metrics, but with the unification expanded from five to six dimensions, the motivation being that if one extra dimension is needed to describe electromagnetism then it is likely that further dimensions will be required for a description of other Standard Model forces – see, for example, Ref. 3. With six dimensions or more there is a vector of vector fields and multiple scalar fields: the generalisation of scalar-tensor gravity to multiple scalar fields has been considered in Refs. 4 and 5, and inflation with multiple scalar fields has been studied in Refs. 6 and 7. In Sec. III the variational principle is applied and the field equations are derived. In Sec. IV the weak-field solution for the scalar fields is found and it is shown how constants may be chosen so that the result is the same as in Brans-Dicke theory. Section V lists the full set of wave solutions – both the four-dimensional ones and those that represent dark matter. Of these solutions, five are consistent with the reflection symmetry, but two dark matter ones are not. In Sec. VI it is shown how these two solutions may be combined into a single solution which is consistent with the reflection symmetry. There are some closing comments in Sec. VII.

---

[a] Electronic mail: trevor.searight@btinternet.com

## II. UNIFICATION IN HIGHER DIMENSIONS

Let indices $a$, $b$, $c$, ... run over 0, 1, 2, 3, 5, 6; let $\mu$, $\nu$, $\rho$, ... run over 0, 1, 2, 3, and let $m$, $n$, ... run over 5, 6. Define a non-degenerate two-index symmetric tensor $\tilde{\gamma}_{ab}$ such that $\tilde{\gamma}_{ab} = \eta_{ab} \equiv$ diag(–1, 1, 1, 1, 1, 1) in flat space. This tensor is not termed the "metric" since it is not what is coupled to matter – the tensor that is coupled to matter is degenerate. However it is useful to define $\tilde{\gamma}_{ab}$ as it is easier to derive the field equations using this quantity. Define vector fields $\varepsilon_a^{(n)}$ such that $\varepsilon_a^{(n)} = \delta_a^n$ in flat space, i.e. such that $\varepsilon_a^{(5)} = (0, 0, 0, 0, 1, 0)$ and $\varepsilon_a^{(6)} = (0, 0, 0, 0, 0, 1)$. Let $\tilde{\gamma}^{ab}$ be the inverse of $\tilde{\gamma}_{ab}$, and define $\varepsilon_{(n)}^a = \tilde{\gamma}^{ab} \varepsilon_b^{(n)}$. Note that the index in brackets is merely used as a counter; it does not imply a coordinate transformation. Finally let the vector fields be normalized so that

$$\tilde{\gamma}^{ab} \varepsilon_a^{(m)} \varepsilon_b^{(n)} = \delta^{mn}. \tag{1}$$

In six dimensions this amounts to three conditions. Thus, when thinking of the six-dimensional fields in (4+2)-dimensional terms (i.e. ignoring the effects of coordinate transformations involve the fifth and sixth coordinates), there are four scalar fields: three from the symmetric tensor $\tilde{\gamma}_{ab}$, two each from the two vector fields $\varepsilon_a^{(5)}$ and $\varepsilon_a^{(6)}$, minus three from the normalization given by equation (1). Generally, in $N$ dimensions, there will be $(N-4)^2$ scalar fields.

The field equations are obtained from the action

$$\int \mathcal{L} \varphi \, dV \tag{2}$$

where $\mathcal{L}$ is the Lagrangian density, $dV$ is the volume element and $\varphi$ is defined by

$$\varphi = \sqrt{|\det(\tilde{\gamma}_{ab})|}. \tag{3}$$

The source-free field equations are required to be invariant under the following transformation, called the eigengauge transformation:

$$\tilde{\gamma}_{ab} \to \tilde{\gamma}_{ab}$$
$$\varepsilon_a^{(n)} \to \varepsilon_a^{(n)} + \partial_a \alpha^{(n)}. \tag{4}$$

(The eigengauge transformation is so called because the vectors $\varepsilon_a^{(n)}$ are eigenvectors of the degenerate metric.) Define the usual connection $\Gamma_{bc}^a$ by

$$\Gamma_{bc}^a = \tfrac{1}{2} \tilde{\gamma}^{ad} (\partial_b \tilde{\gamma}_{cd} + \partial_c \tilde{\gamma}_{bd} - \partial_d \tilde{\gamma}_{bc}), \tag{5}$$

and antisymmetric tensors $E_{ab}^{(n)}$ by

$$E_{ab}^{(n)} = \partial_a \varepsilon_b^{(n)} - \partial_b \varepsilon_a^{(n)}; \tag{6}$$

$\Gamma_{bc}^a$ and $E_{ab}^{(n)}$ are eigengauge invariant; $\varphi$ is also eigengauge invariant. The eigengauge invariant Lagrangian density is then

$$\mathcal{L} = \frac{c^3}{16\pi G_0}\left(R - 2\Lambda - \tfrac{1}{2}\sum_{m,n}\omega_{mn}\widetilde{\gamma}^{ab}\widetilde{\gamma}^{cd} E_{ac}^{(m)} E_{bd}^{(n)}\right) \quad (7)$$

where $R \equiv \widetilde{\gamma}^{bd}(\partial_a\Gamma_{bd}^a - \partial_b\Gamma_{ad}^a + \Gamma_{bd}^e\Gamma_{ae}^a - \Gamma_{ad}^e\Gamma_{be}^a)$ is the Ricci scalar, $\omega_{mn}$ is a symmetric matrix of dimensionless constants, $\Lambda$ is the cosmological constant, and $G_0$ is a constant with the same dimensions as Newton's constant of gravitation $G$.

Now turn to the equations of motion. The key to the unification is to arrange it so that

$$(\varepsilon_a^{(m)} u^a)(\varepsilon_{(n)}^b p_b) = 0 \quad (8)$$

for all $m$ and $n$, where $u^a$ is the $N$-velocity of the particle and $p_a$ is the $N$-momentum. Equation (8) is needed in order that there are no unwanted sources of the type $q^2/m$, as described in Ref. 1. One way to achieve this is to use a Lagrangian

$$L = \tfrac{1}{2} mc\gamma_{ab} u^a u^b \quad (9)$$

with degenerate metric $\gamma_{ab}$ defined by

$$\gamma_{ab} = \widetilde{\gamma}_{ab} - \sum_n \varepsilon_a^{(n)}\varepsilon_b^{(n)} . \quad (10)$$

The vectors $\varepsilon_{(n)}^a$ are eigenvectors of $\gamma_{ab}$ with eigenvalue zero, i.e.

$$\gamma_{ab}\varepsilon_{(n)}^b = 0, \quad (11)$$

and since $p_a = mc\gamma_{ab}u^b$ it follows that $\varepsilon_{(n)}^a p_a = 0$ for all $n$, and so equation (8) is satisfied. Another approach is to use the Hamiltonian

$$H = \frac{1}{2mc}\gamma^{ab} p_a p_b \quad (12)$$

with degenerate metric $\gamma^{ab}$ defined by

$$\gamma^{ab} = \widetilde{\gamma}^{ab} - \sum_n \varepsilon_{(n)}^a \varepsilon_{(n)}^b . \quad (13)$$

The vectors $\varepsilon_a^{(n)}$ are eigenvectors of $\gamma^{ab}$ with eigenvalue zero, i.e.

$$\gamma^{ab}\varepsilon_b^{(n)} = 0, \quad (14)$$

and since $u^a = \gamma^{ab} p_b / mc$ it follows that $\varepsilon_a^{(n)} u^a = 0$ for all $n$, and so equation (8) is satisfied once again. Thus the degenerate metric can be thought of as the thing which "delivers" equation (8). A degenerate metric does not have an inverse, however $\gamma_{ab}$ and $\gamma^{ab}$ obey

$$\gamma_{ab}\gamma^{bc}\gamma_{cd} = \gamma_{ad} \quad (15)$$

and

$$\gamma^{ab}\gamma_{bc}\gamma^{cd} = \gamma^{ad}. \tag{16}$$

The equations of motion are obtained from the Lagrangian or the Hamiltonian in the usual way – provided that the same Lagrangian or Hamiltonian is used for all particles – however in both cases some further equations are obtained by making use of the eigengauge invariance. Because the Lagrangian density (7) is eigengauge invariant, when one applies the eigengauge transformation to the combination of field and matter, only the matter terms change, and therefore new equations result for matter (so that, paradoxically, some further equations of motion are obtained by varying the field). In the case of the Lagrangian (9) one finds

$$\varepsilon_a^{(n)} u^a = \text{constant} \tag{17}$$

for all $n$, while in the case of the Hamiltonian (12) one finds

$$\varepsilon_{(n)}^a p_a = \text{constant}. \tag{18}$$

The Lagrangian (7) is a unified Lagrangian because there is a reflection symmetry between the two sets of vector fields $\varepsilon_a^{(n)}$ and $\varepsilon_{(n)}^a$, so that charges may either described by the $N$-momentum, or by the $N$-velocity. For consider a static, weak field where

$$\widetilde{\gamma}_{ab} = \begin{pmatrix} -1 & 0 & \psi+\chi & \widehat{\psi}+\widehat{\chi} \\ 0 & \delta_{ij} & 0 & 0 \\ \psi+\chi & 0 & 1 & 0 \\ \widehat{\psi}+\widehat{\chi} & 0 & 0 & 1 \end{pmatrix} \qquad \varepsilon_a^{(5)} = \begin{pmatrix} \psi \\ 0 \\ 1 \\ 0 \end{pmatrix} \qquad \varepsilon_a^{(6)} = \begin{pmatrix} \widehat{\psi} \\ 0 \\ 0 \\ 1 \end{pmatrix}$$

(ignoring second order terms). Then the Ricci scalar contains

$$\tfrac{1}{2}(\partial_i\psi)(\partial_i\psi) + (\partial_i\psi)(\partial_i\chi) + \tfrac{1}{2}(\partial_i\chi)(\partial_i\chi)$$
$$+ \tfrac{1}{2}(\partial_i\widehat{\psi})(\partial_i\widehat{\psi}) + (\partial_i\widehat{\psi})(\partial_i\widehat{\chi}) + \tfrac{1}{2}(\partial_i\widehat{\chi})(\partial_i\widehat{\chi})$$

while the scalar $-\tfrac{1}{2}\sum_{m,n}\omega_{mn}\widetilde{\gamma}^{ab}\widetilde{\gamma}^{cd}E_{ac}^{(m)}E_{bd}^{(n)}$ contains

$$\omega_{55}(\partial_i\psi)(\partial_i\psi) + 2\omega_{56}(\partial_i\psi)(\partial_i\widehat{\psi}) + \omega_{66}(\partial_i\widehat{\psi})(\partial_i\widehat{\psi}).$$

The quadratic terms in the two scalars must be added together so as to obtain Maxwell's equations for both sets of vector fields. With the exception of the cosmological constant, the terms in the Lagrangian (7) have to be combined in this linear fashion. In the case where charges are described by the $N$-momentum, it is $\varepsilon_a^{(n)}$ which is coupled to matter – indirectly through the degenerate metric $\gamma^{ab}$ in the Hamiltonian (12); the two charges are $\varepsilon_{(5)}^a p_a$ and $\varepsilon_{(6)}^a p_a$. In the case where charges are described by the $N$-velocity, it is $\varepsilon_{(n)}^a$ which is coupled to matter – indirectly through the degenerate metric $\gamma_{ab}$ in the Lagrangian (9); the two charges are $\varepsilon_a^{(5)} u^a$ and $\varepsilon_a^{(6)} u^a$.

The electromagnetic terms in the Lagrangian (7) may be summarized in matrix notation as

$$\tfrac{1}{2}(\psi \ \ \widehat{\psi} \ \ \chi \ \ \widehat{\chi}) \begin{pmatrix} 1+2\omega_{55} & 2\omega_{56} & 1 & 0 \\ 2\omega_{56} & 1+2\omega_{66} & 0 & 1 \\ 1 & 0 & 1 & 0 \\ 0 & 1 & 0 & 1 \end{pmatrix} \begin{pmatrix} \psi \\ \widehat{\psi} \\ \chi \\ \widehat{\chi} \end{pmatrix}, \tag{19}$$

or more succinctly as

$$\tfrac{1}{2}(\Psi \ \ X)\begin{pmatrix} I+2\Omega & I \\ I & I \end{pmatrix}\begin{pmatrix} \Psi \\ X \end{pmatrix} \tag{20}$$

where

$$\Omega = \begin{pmatrix} \omega_{55} & \omega_{56} \\ \omega_{56} & \omega_{66} \end{pmatrix} \qquad \Psi = \begin{pmatrix} \psi \\ \widehat{\psi} \end{pmatrix} \qquad X = \begin{pmatrix} \chi \\ \widehat{\chi} \end{pmatrix}.$$

Under the reflection symmetry, $\chi$ and $\widehat{\chi}$ are replaced by linear combinations of $\psi$ and $\widehat{\psi}$, and $\psi$ and $\widehat{\psi}$ are replaced by linear combinations of $\chi$ and $\widehat{\chi}$, while preserving the expression (19). Thus $X \to A\Psi$ for some 2×2 matrix $A$, while $\Psi \to A^{-1}X$. Note that $A^{-1}$ is needed so that if the reflection is applied twice the four fields return to their original values. The matrix $A$ must therefore satisfy

$$\begin{pmatrix} 0 & A^T \\ (A^{-1})^T & 0 \end{pmatrix}\begin{pmatrix} I+2\Omega & I \\ I & I \end{pmatrix}\begin{pmatrix} 0 & A^{-1} \\ A & 0 \end{pmatrix} = \begin{pmatrix} I+2\Omega & I \\ I & I \end{pmatrix}. \tag{21}$$

Multiplying out the expression on the left-hand side of equation (21) one finds that $A = A^T$, i.e. $A$ is a symmetric matrix, and $A^2 = I + 2\Omega$. Using the formula for the square root of a 2×2 matrix[8] one obtains

$$A = \frac{(I+2\Omega) + I\sqrt{\det(I+2\Omega)}}{\sqrt{\operatorname{tr}(I+2\Omega) + 2\sqrt{\det(I+2\Omega)}}}. \tag{22}$$

The important point to note is that all three constants $\omega_{55}, \omega_{56}$ and $\omega_{66}$ are inextricably linked in equation (22), meaning that the reflection symmetry unifies all the terms in the Lagrangian (7).

### III. THE FIELD EQUATIONS

The field equations are

$$\partial_a \frac{\partial(\mathcal{L}\varphi)}{\partial_a \phi_\alpha} = \frac{\partial(\mathcal{L}\varphi)}{\partial \phi_\alpha} \tag{23}$$

for fields $\phi_\alpha$, where $\mathcal{L} = \mathcal{L}_{field} + \mathcal{L}_{matter}$. Choosing to vary the degenerate metrics and their eigenvectors, introduce a new term

$$\beta_b^a \left( \delta_a^b - \gamma^{bc}\gamma_{ca} - \sum_n \varepsilon_{(n)}^b \varepsilon_a^{(n)} \right) + \sum_{m,n} \zeta_n^m \left( \delta_m^n - \varepsilon_{(m)}^a \varepsilon_a^{(n)} \right) \tag{24}$$

to the Lagrangian, where the $\beta_b^a$ and $\zeta_n^m$ are Lagrange multipliers, so that $\gamma_{ab}$, $\varepsilon_a^{(n)}$, $\gamma^{ab}$ and $\varepsilon_{(n)}^a$ can be treated as independent when being varied. The Lagrangian multipliers can then be eliminated from the resulting equations leaving just equations for the field. The raw equations are

$$\partial_a \frac{\partial(\mathcal{L}\varphi)}{\partial_a \gamma_{bc}} = \frac{\partial(\mathcal{L}\varphi)}{\partial \gamma_{bc}} - \tfrac{1}{2}\varphi(\beta_a^b \gamma^{ac} + \beta_a^c \gamma^{ab}) \qquad (25)$$

$$\partial_a \frac{\partial(\mathcal{L}\varphi)}{\partial_a \gamma^{bc}} = \frac{\partial(\mathcal{L}\varphi)}{\partial \gamma^{bc}} - \tfrac{1}{2}\varphi(\beta_b^a \gamma_{ac} + \beta_c^a \gamma_{ab}) \qquad (26)$$

$$\partial_a \frac{\partial(\mathcal{L}\varphi)}{\partial_a \varepsilon_b^{(n)}} = \frac{\partial(\mathcal{L}\varphi)}{\partial \varepsilon_b^{(n)}} - \varphi \beta_a^b \varepsilon_{(n)}^a - \varphi \sum_p \zeta_n^p \varepsilon_{(p)}^b \qquad (27)$$

and

$$\partial_a \frac{\partial(\mathcal{L}\varphi)}{\partial_a \varepsilon_{(n)}^b} = \frac{\partial(\mathcal{L}\varphi)}{\partial \varepsilon_{(n)}^b} - \varphi \beta_b^a \varepsilon_a^{(n)} - \varphi \sum_p \zeta_p^n \varepsilon_b^{(p)} \ . \qquad (28)$$

Define $h_b^a = \gamma^{ac} \gamma_{cb}$, a tensor. Then equation (25) contracted with $h_b^d h_c^e$ minus equation (26) contracted with $\gamma^{bd} \gamma^{ce}$ gives

$$\partial_a \left( \frac{\partial(\mathcal{L}\varphi)}{\partial_a \gamma_{bc}} \right) h_b^d h_c^e - \partial_a \left( \frac{\partial(\mathcal{L}\varphi)}{\partial_a \gamma^{bc}} \right) \gamma^{bd} \gamma^{ce} = \frac{\partial(\mathcal{L}\varphi)}{\partial \gamma_{bc}} h_b^d h_c^e - \frac{\partial(\mathcal{L}\varphi)}{\partial \gamma^{bc}} \gamma^{bd} \gamma^{ce} \ ; \qquad (29)$$

equation (25) contracted with $\varepsilon_b^{(n)} h_c^e$ minus $\tfrac{1}{2}$ times equation (28) contracted with $\gamma^{be}$ gives

$$\partial_a \left( \frac{\partial(\mathcal{L}\varphi)}{\partial_a \gamma_{bc}} \right) \varepsilon_b^{(n)} h_c^e - \tfrac{1}{2} \partial_a \left( \frac{\partial(\mathcal{L}\varphi)}{\partial_a \varepsilon_{(n)}^b} \right) \gamma^{be} = \frac{\partial(\mathcal{L}\varphi)}{\partial \gamma_{bc}} \varepsilon_b^{(n)} h_c^e - \tfrac{1}{2} \frac{\partial(\mathcal{L}\varphi)}{\partial \varepsilon_{(n)}^b} \gamma^{be} \ ; \qquad (30)$$

equation (26) contracted with $\varepsilon_{(n)}^b h_e^c$ minus $\tfrac{1}{2}$ times equation (27) contracted with $\gamma_{be}$ gives

$$\partial_a \left( \frac{\partial(\mathcal{L}\varphi)}{\partial_a \gamma^{bc}} \right) \varepsilon_{(n)}^b h_e^c - \tfrac{1}{2} \partial_a \left( \frac{\partial(\mathcal{L}\varphi)}{\partial_a \varepsilon_b^{(n)}} \right) \gamma_{be} = \frac{\partial(\mathcal{L}\varphi)}{\partial \gamma^{bc}} \varepsilon_{(n)}^b h_e^c - \tfrac{1}{2} \frac{\partial(\mathcal{L}\varphi)}{\partial \varepsilon_b^{(n)}} \gamma_{be} \ ; \qquad (31)$$

while equation (27) contracted with $\varepsilon_b^{(m)}$ minus equation (28) with $n=m$ contracted with $\varepsilon_{(n)}^b$ gives

$$\partial_a \left( \frac{\partial(\mathcal{L}\varphi)}{\partial_a \varepsilon_b^{(n)}} \right) \varepsilon_b^{(m)} - \partial_a \left( \frac{\partial(\mathcal{L}\varphi)}{\partial_a \varepsilon_{(m)}^b} \right) \varepsilon_{(n)}^b = \frac{\partial(\mathcal{L}\varphi)}{\partial \varepsilon_b^{(n)}} \varepsilon_b^{(m)} - \frac{\partial(\mathcal{L}\varphi)}{\partial \varepsilon_{(m)}^b} \varepsilon_{(n)}^b \ . \qquad (32)$$

From equation (29) one finds tensor equation

$$\tfrac{1}{2}(R_b^a h_a^e \gamma^{bf} + R_b^a h_a^f \gamma^{be}) - \sum_{m,n} \omega_{mn} \gamma^{ae} \gamma^{bf} \tilde{\gamma}^{cd} E_{ac}^{(m)} E_{bd}^{(n)}$$

$$- \tfrac{1}{2} \gamma^{ef} (R - 2\Lambda - \tfrac{1}{2} \sum_{m,n} \omega_{mn} \tilde{\gamma}^{ab} \tilde{\gamma}^{cd} E_{ac}^{(m)} E_{bd}^{(n)}) = \frac{16\pi G_0}{c^3} \sum_{\text{particles}} \frac{1}{2mc} p_a p_b \gamma^{ae} \gamma^{bf} \rho \ , \qquad (33)$$

where $R_c^a \equiv \tilde{\gamma}^{bd}(\partial_c \Gamma_{bd}^a - \partial_b \Gamma_{cd}^a + \Gamma_{bd}^e \Gamma_{ce}^a - \Gamma_{cd}^e \Gamma_{be}^a)$ is the Ricci tensor and $\rho$ is the particle density. Contracting equation (33) with $\gamma_{ef}$ gives scalar equation

$$R - 4\Lambda - \varphi^{-1}\partial_a\left(\varphi\tilde{\gamma}^{ab}\sum_n E_{bc}^{(n)}\varepsilon_{(n)}^c\right) - \tfrac{1}{2}\sum_n \varepsilon_{(n)}^a(\partial_a E_{(n)}) - \tfrac{1}{4}\tilde{\gamma}_{ab}\tilde{\gamma}_{cd}\sum_n E_{(n)}^{ac}E_{(n)}^{bd}$$
$$+ \tfrac{1}{4}\tilde{\gamma}^{ab}\tilde{\gamma}^{cd}\sum_n E_{ac}^{(n)}E_{bd}^{(n)} - \sum_{m,n}\sum_p \omega_{mn}\tilde{\gamma}^{cd}\varepsilon_{(p)}^a E_{ac}^{(m)}\varepsilon_{(p)}^b E_{bd}^{(n)} = \frac{8\pi G_0}{c^3}\sum_{\text{particles}} mc\rho, \qquad (34)$$

where the $E_{(n)}^{ab}$ are symmetric tensors defined by

$$E_{(n)}^{ab} = \tilde{\gamma}^{ac}(\partial_c \varepsilon_{(n)}^b) + \tilde{\gamma}^{bc}(\partial_c \varepsilon_{(n)}^a) - \varepsilon_{(n)}^c(\partial_c \tilde{\gamma}^{ab}) \qquad (35)$$

and $E_{(n)} \equiv \tilde{\gamma}_{ab}E_{(n)}^{ab}$. From equation (32) one finds scalar equations

$$\varphi^{-1}\partial_a(\varphi\tilde{\gamma}^{ab}E_{bc}^{(m)}\varepsilon_{(n)}^c) + \varphi^{-1}\partial_a(\varphi\tilde{\gamma}^{ab}E_{bc}^{(n)}\varepsilon_{(m)}^c) + \tfrac{1}{2}\varepsilon_{(m)}^a(\partial_a E_{(n)}) + \tfrac{1}{2}\varepsilon_{(n)}^a(\partial_a E_{(m)})$$
$$+ \tfrac{1}{2}\tilde{\gamma}_{ab}\tilde{\gamma}_{cd}E_{(m)}^{ac}E_{(n)}^{bd} - \tfrac{1}{2}\tilde{\gamma}^{ab}\tilde{\gamma}^{cd}E_{ac}^{(m)}E_{bd}^{(n)}$$
$$+ 2\varphi^{-1}\partial_a\left(\varphi\sum_p \omega_{np}\tilde{\gamma}^{ab}E_{bd}^{(p)}\varepsilon_{(m)}^d\right) - \sum_p \omega_{np}\tilde{\gamma}^{ab}\tilde{\gamma}^{cd}E_{ac}^{(m)}E_{bd}^{(p)} + 2\sum_{p,q}\omega_{pq}\varepsilon_{(m)}^a\varepsilon_{(n)}^b\tilde{\gamma}^{cd}E_{ac}^{(p)}E_{bd}^{(q)}$$
$$+ \delta_{mn}\left(R - 2\Lambda - \tfrac{1}{2}\sum_{p,q}\omega_{pq}\tilde{\gamma}^{ab}\tilde{\gamma}^{cd}E_{ac}^{(p)}E_{bd}^{(q)}\right) = 0 \qquad (36)$$

which amounts to four equations in six dimensions; with equation (34) there is a total of five scalar equations. From equation (30) one finds vector equation

$$\tfrac{1}{2}(\nabla_a E_{(n)}^{ab})h_b^e - \tfrac{1}{2}(\partial_a E_{(n)})\gamma^{ae} - \tfrac{1}{2}(\nabla_a E_{bc}^{(n)})\tilde{\gamma}^{ab}\gamma^{ce}$$
$$- \sum_{p,q}\omega_{pq}\varepsilon_{(n)}^c\tilde{\gamma}^{ab}E_{ac}^{(p)}E_{bd}^{(q)}\gamma^{de} = 0 \qquad (37)$$

assuming the equations of motion are derived from the Hamiltonian (12); in the case where the equations of motion are derived from the Lagrangian (9) the right-hand side is

$$\frac{16\pi G_0}{c^3}\sum_{\text{particles}} \tfrac{1}{2}mc(\varepsilon_a^{(n)}u^a)u^b h_b^e\rho.$$

From equation (31) one finds vector equation

$$\tfrac{1}{2}(\nabla_a E_{(n)}^{ab})h_b^e - \tfrac{1}{2}(\partial_a E_{(n)})\gamma^{ae} - \tfrac{1}{2}(\nabla_a E_{bc}^{(n)})\tilde{\gamma}^{ab}\gamma^{ce} - \nabla_a\left(\sum_p \omega_{np}E_{bc}^{(p)}\right)\tilde{\gamma}^{ab}\gamma^{ce}$$
$$- \sum_{p,q}\omega_{pq}\varepsilon_{(n)}^c\tilde{\gamma}^{ab}E_{ac}^{(p)}E_{bd}^{(q)}\gamma^{de} = \frac{16\pi G_0}{c^3}\sum_{\text{particles}} \frac{1}{2mc}(\varepsilon_{(n)}^a p_a)p_b\gamma^{be}\rho \qquad (38)$$

assuming the equations of motion are derived from the Hamiltonian (12); in the case where the equations of motion are derived from the Lagrangian (9) the right-hand side is zero. Note that the left-hand side of equation (38) minus the left-hand side of equation (37) is

$$-\nabla_a\left(\sum_p \omega_{np} E^{(p)}_{bc}\right)\tilde{\gamma}^{ab}\gamma^{ce}.$$

This expression is the matrix $\Omega$ multiplied by a vector of second derivatives. In order that this expression may be inverted to solve for the field it follows that $\det(\Omega)$ must be non-zero.

## IV. UNDERSTANDING THE SCALAR FIELDS

In order to understand the effect of the scalar fields, for simplicity let the 4-vector fields be zero, i.e.

$$\gamma_{ab} = \begin{pmatrix} g_{\mu\nu} & 0 & 0 \\ 0 & 0 & 0 \\ 0 & 0 & 0 \end{pmatrix} \quad \gamma^{ab} = \begin{pmatrix} g^{\mu\nu} & 0 & 0 \\ 0 & 0 & 0 \\ 0 & 0 & 0 \end{pmatrix},$$

let the cosmological constant be zero, and assume weak fields. Note that at infinity $\varepsilon^{(5)}_5$ and $\varepsilon^{(6)}_6$ are both 1 while $\varepsilon^{(5)}_6$ and $\varepsilon^{(6)}_5$ are both zero. Also assume that fields are independent of both the fifth and sixth coordinates, i.e. $\partial_5(\text{fields}) = 0$ and $\partial_6(\text{fields}) = 0$.

The Ricci scalar can be eliminated from equation (36) using equation (34) to obtain four scalar equations which may be written in matrix form as

$$\begin{pmatrix} 2\omega_{55}+3 & 2\omega_{56} & 1 & 0 \\ 2\omega_{56} & 2\omega_{66}+1 & 0 & 1 \\ 1 & 0 & 2\omega_{66}+3 & 2\omega_{56} \\ 0 & 1 & 2\omega_{56} & 2\omega_{55}+1 \end{pmatrix} \begin{pmatrix} \varphi^{-1}\partial_\mu(\varphi g^{\mu\nu}(\partial_\nu \varepsilon^{(5)}_5)) \\ \varphi^{-1}\partial_\mu(\varphi g^{\mu\nu}(\partial_\nu \varepsilon^{(6)}_5)) \\ \varphi^{-1}\partial_\mu(\varphi g^{\mu\nu}(\partial_\nu \varepsilon^{(6)}_6)) \\ \varphi^{-1}\partial_\mu(\varphi g^{\mu\nu}(\partial_\nu \varepsilon^{(5)}_6)) \end{pmatrix}$$

$$= -\frac{8\pi G_0}{c^3}\sum_{\text{particles}} mc\rho \begin{pmatrix} 1 \\ 0 \\ 1 \\ 0 \end{pmatrix}. \quad (39)$$

The simplest case to examine is $\Omega = \omega I$, where $\omega$ is a dimensionless constant. In this case there is a global O(2) symmetry on the vector of vector fields. This can be seen easily in the Lagrangian density (7) since $\tilde{\gamma}_{ab}$ obeys the symmetry and consequently so does the Ricci scalar. At first sight the 4×4 matrix on the left-hand side of equation (39) does not obey the symmetry, however this is simply a consequence of having arranged $\varepsilon^{(n)}_a = \delta_{na}$ at infinity: a coordinate transformation that rotates between the fifth and sixth coordinates must be applied to preserve the matrix. Note that, as in Brans-Dicke theory, the constant $\omega$ is a constant of gravitation. The strengths of the vector fields is governed by the magnitudes of $\varepsilon^{(n)}_a u^a$ or $\varepsilon^a_{(n)} p_a$ in which there are embedded constants. In the case of electromagnetism $\omega$ is absorbed by the embedded constant, and the permeability constant $\mu_0$ replaces it.

If $\Omega = \omega I$ equation (39) can be inverted to give

$$\varphi^{-1}\partial_\mu(\varphi g^{\mu\nu}(\partial_\nu \varepsilon^{(5)}_5)) = -\frac{4\pi G_0}{(\omega+2)c^3}\sum_{\text{particles}} mc\rho \quad (40)$$

and

$$\varphi^{-1}\partial_\mu(\varphi g^{\mu\nu}(\partial_\nu \varepsilon_5^{(6)})) = 0 \tag{41}$$

and similarly for $\varepsilon_6^{(6)}$ and $\varepsilon_6^{(5)}$. Thus the spherically symmetric solution is

$$\varepsilon_5^{(5)} = \varepsilon_6^{(6)} = 1 + \frac{G_0 m}{(\omega+2)c^2 r} \tag{42}$$

and

$$\varepsilon_5^{(6)} = \varepsilon_6^{(5)} = 0. \tag{43}$$

For a weak field $\varphi = \varepsilon_5^{(5)} \varepsilon_6^{(6)} \sqrt{|\det g_{\mu\nu}|}$ and equation (33) can be solved from this in the same way as in Brans-Dicke theory to give

$$g_{00} = -1 + \frac{2G_0 m}{c^2 r}\left(1 + \frac{1}{(\omega+2)}\right) \tag{44}$$

$$g_{ii} = 1 + \frac{2G_0 m}{c^2 r}\left(1 - \frac{1}{(\omega+2)}\right) \tag{45}$$

and $g_{ij} = 0$ for $i \neq j$. The constant $G_0$ is determined by the strength of gravity, so that $g_{00} = -1 + 2Gm/c^2 r$. Thus

$$G_0 = G\left(\frac{\omega+2}{\omega+3}\right). \tag{46}$$

Substituting this into equation (45) gives

$$g_{ii} = 1 + \frac{2Gm}{c^2 r}\left(\frac{\omega+1}{\omega+3}\right). \tag{47}$$

The equivalent result from Brans-Dicke theory is

$$g_{ii} = 1 + \frac{2Gm}{c^2 r}\left(\frac{\omega_{BD}+1}{\omega_{BD}+2}\right). \tag{48}$$

Thus the metric behaves differently with respect to the constant $\omega$ in six dimensions than it does with five. Equating the right-hand side of equation (47) with the right-hand side of equation (48) gives

$$\omega = 2\omega_{BD} + 1 \tag{49}$$

and then the solution for $g_{\mu\nu}$ is the same as in Brans-Dicke theory. There are four values that $\omega$ cannot take. The determinant of the matrix in equation (39) is $16\omega(\omega+2)(\omega+1)^2$ and so in order to be able to invert the matrix $\omega$ cannot be equal to 0, –1 or –2. Additionally, equation (47) implies that $\omega$ cannot be equal to –3. Thus $\omega_{BD}$ cannot take the values $-\frac{1}{2}$, –1, $-\frac{3}{2}$, –2. In a general $N$ dimensions the Brans-Dicke results are recovered with

$$\omega = (N-4)\omega_{BD} + (N-5). \tag{50}$$

The deviations from general relativity predicted by the Brans-Dicke theory can be tested for experimentally, and negative results can be used to put a bound on the value of $\omega_{BD}$. A radar signal sent to a planet or satellite which passes close to the Sun experiences a non-Newtonian delay in its travel-time: results from the Cassini spacecraft while on its way to Saturn put $\omega_{BD}$ at greater than 40 000 (Ref. 9). Gravitational wave detection from inspiralling compact binary systems with the proposed Einstein Telescope may provide a way to increase the level of the constraint[10]. The existence of a bound for $\omega_{BD}$, however large, does not disprove the degenerate theory (or Brans-Dicke theory). And, as noted above, $\omega$ is a constant of gravitation, and its value has no bearing on the strength of electromagnetism.

## V. THE WAVE SOLUTIONS

### A. The Field Equations in "(4+1)-Dimensional" Notation

In Secs. V and VI work in five dimensions, dropping the bracketed indices $m$, $n$ so that $\varepsilon_a^{(n)}$ is written $\varepsilon_a$ (a single vector) and $\varepsilon_{(n)}^a$ is written $\varepsilon^a$; the matrix $\Omega$ is replaced by the constant $\omega$. To find the wave solutions to the field equations work in the following "(4+1)-dimensional" notation

$$\gamma^{ab} = \begin{pmatrix} g^{\mu\nu} & -g^{\mu\rho}\varepsilon_\rho\varepsilon_5^{-1} \\ -g^{\nu\sigma}\varepsilon_\sigma\varepsilon_5^{-1} & (g^{\rho\sigma}\varepsilon_\rho\varepsilon_\sigma)\varepsilon_5^{-2} \end{pmatrix}$$

$$\varepsilon_a = (\varepsilon_\mu, \quad \varepsilon_5)$$

$$\varepsilon^a = (\varepsilon^\mu, \quad \varepsilon_5^{-1} - \varepsilon_\rho\varepsilon^\rho\varepsilon_5^{-1})$$

(51)

$$\gamma_{ab} = \begin{pmatrix} g_{\mu\nu} - g_{\mu\rho}\varepsilon^\rho\varepsilon_\nu - g_{\nu\rho}\varepsilon^\rho\varepsilon_\mu + (g_{\rho\sigma}\varepsilon^\rho\varepsilon^\sigma)\varepsilon_\mu\varepsilon_\nu & -g_{\mu\rho}\varepsilon^\rho\varepsilon_5 + (g_{\rho\sigma}\varepsilon^\rho\varepsilon^\sigma)\varepsilon_\mu\varepsilon_5 \\ -g_{\nu\sigma}\varepsilon^\sigma\varepsilon_5 + (g_{\rho\sigma}\varepsilon^\rho\varepsilon^\sigma)\varepsilon_\nu\varepsilon_5 & (g_{\rho\sigma}\varepsilon^\rho\varepsilon^\sigma)(\varepsilon_5)^2 \end{pmatrix}$$

and assume weak fields and cosmological constant zero. Then equation (33) becomes

$$\tfrac{1}{2} g^{\lambda\mu} g^{\nu\sigma} g^{\rho\tau}(\partial_\nu(\partial_\lambda g_{\mu\rho}) - \partial_\nu(\partial_\rho g_{\mu\lambda}) + \partial_\lambda(\partial_\rho g_{\mu\nu}) - \partial_\lambda(\partial_\mu g_{\rho\nu})) - g^{\nu\sigma} g^{\rho\tau}(\partial_\nu(\partial_\rho \varepsilon_5))$$
$$+ \tfrac{1}{2} g^{\nu\sigma} g^{\rho\tau}(\partial_5(\partial_\nu \varepsilon_\rho)) + \tfrac{1}{2} g^{\nu\sigma} g^{\rho\tau}(\partial_5(\partial_\rho \varepsilon_\nu)) - \tfrac{1}{2} g^{\nu\sigma}(\partial_5(\partial_\nu \varepsilon^\tau)) - \tfrac{1}{2} g^{\nu\tau}(\partial_5(\partial_\nu \varepsilon^\sigma))$$
$$- \tfrac{1}{2} g^{\nu\sigma} g^{\rho\tau}(\partial_5(\partial_5 g_{\nu\rho})) - \tfrac{1}{2} g^{\sigma\tau} g^{\lambda\mu} g^{\nu\rho}(\partial_\lambda(\partial_\nu g_{\mu\rho} - \partial_\mu g_{\nu\rho}))$$
$$+ g^{\sigma\tau} g^{\mu\nu}(\partial_\mu(\partial_\nu \varepsilon_5 - \partial_5 \varepsilon_\nu)) + g^{\sigma\tau}(\partial_5(\partial_\mu \varepsilon^\mu)) + \tfrac{1}{2} g^{\sigma\tau}(\partial_5(\partial_5 g_{\mu\nu})) g^{\mu\nu} = 0; \qquad (52)$$

combining equations (34) and (36) gives

$$(2\omega + 3) g^{\mu\nu}(\partial_\mu(\partial_\nu \varepsilon_5 - \partial_5 \varepsilon_\nu)) + 3(\partial_5(\partial_\mu \varepsilon^\mu)) + \tfrac{3}{2}(\partial_5(\partial_5 g_{\mu\nu})) g^{\mu\nu} = 0, \qquad (53)$$

and combining equations (37) and (38) gives

$$(\partial_\mu(\partial_\nu \varepsilon_\rho - \partial_\rho \varepsilon_\nu)) g^{\mu\nu} g^{\rho\sigma} + (\partial_5(\partial_5 \varepsilon_\rho - \partial_\rho \varepsilon_5)) g^{\rho\sigma} = 0 \qquad (54)$$

and

$$(\partial_\mu(\partial_\nu\varepsilon^\sigma))g^{\mu\nu} - (\partial_\rho(\partial_\mu\varepsilon^\mu))g^{\rho\sigma} + (\partial_5(\partial_5\varepsilon_\rho - \partial_\rho\varepsilon_5))g^{\rho\sigma}$$
$$- (\partial_\rho(\partial_5 g_{\mu\nu}))g^{\mu\nu}g^{\rho\sigma} - (\partial_5(\partial_\mu g^{\mu\sigma})) = 0. \tag{55}$$

Since the equations for the weak field are linear, the simplest way to solve them is to find a set of individual solutions which can then be combined into a general solution. These include familiar four-dimensional solutions for which no special comment is required; however, they are given for completeness.

### B. Covariant Photon

The first solution is the photon

$$\varepsilon_\mu = a_\mu \exp(ik_\rho x^\rho)$$
$$\varepsilon_5 = 1$$
$$\varepsilon^\mu = 0 \tag{56}$$
$$g_{\mu\nu} = \eta_{\mu\nu},$$

where $g^{\mu\nu}a_\mu k_\nu = 0$, $g^{\mu\nu}k_\mu k_\nu = 0$ and $\eta_{\mu\nu}$ is the Minkowski metric.

### C. Contravariant Photon

The second solution is also a photon:

$$\varepsilon_\mu = 0$$
$$\varepsilon_5 = 1$$
$$\varepsilon^\mu = a^\mu \exp(ik_\rho x^\rho) \tag{57}$$
$$g_{\mu\nu} = \eta_{\mu\nu},$$

where $a^\mu k_\mu = 0$ and $g^{\mu\nu}k_\mu k_\nu = 0$.

### D. Scalar Particle

The third solution is a wave of the scalar field

$$\varepsilon_\mu = 0$$
$$\varepsilon_5 = 1 + a_5 \exp(ik_\rho x^\rho)$$
$$\varepsilon^\mu = 0 \tag{58}$$
$$g_{\mu\nu} = \eta_{\mu\nu} + a_{\mu\nu} \exp(ik_\rho x^\rho),$$

where $g^{\mu\nu}k_\mu k_\nu = 0$ and

$$\tfrac{1}{2} g^{\mu\nu}(k_\rho k_\mu a_{\nu\sigma} + k_\sigma k_\mu a_{\nu\rho}) - \tfrac{1}{2} g^{\mu\nu} a_{\mu\nu} k_\rho k_\sigma - a_5 k_\rho k_\sigma = 0. \tag{59}$$

Equation (59) can be solved with $a_{\mu\nu} = a_\mu a_\nu$ if $g^{\mu\nu} a_\mu k_\nu = 0$ and $\tfrac{1}{2} g^{\mu\nu} a_\mu a_\nu + a_5 = 0$.

### E. Graviton

The fourth solution to the field equations is the graviton

$$\varepsilon_\mu = 0$$

$$\varepsilon_5 = 1 \tag{60}$$

$$\varepsilon^\mu = 0$$

$$g_{\mu\nu} = \eta_{\mu\nu} + a_{\mu\nu} \exp(ik_\rho x^\rho),$$

where

$$g^{\lambda\mu} g^{\nu\sigma} g^{\rho\tau}(k_\lambda k_\nu a_{\mu\rho} + k_\lambda k_\rho a_{\mu\nu} - k_\nu k_\rho a_{\lambda\mu} - k_\lambda k_\mu a_{\nu\rho}) = 0. \tag{61}$$

### F. Massive Vector Particle

Now turn to solutions which allow for the wave behaviour to extend to the fifth dimension. The first of these is the solution which was given in Ref. 1 that combined two massive photons:

$$\varepsilon_\mu = a_\mu \exp(ik_\rho x^\rho + ik_5 x^5)$$

$$\varepsilon_5 = 1 \tag{62}$$

$$\varepsilon^\mu = g^{\mu\nu} a_\nu \exp(ik_\rho x^\rho + ik_5 x^5)$$

$$g_{\mu\nu} = \eta_{\mu\nu},$$

where $g^{\mu\nu} a_\mu k_\nu = 0$ and $g^{\mu\nu} k_\mu k_\nu + (k_5)^2 = 0$. In Ref. 1 this solution was interpreted as dark matter, since it acts as a source of gravity but does not interact with ordinary matter. Ordinary matter is spread out evenly across the fifth dimension, so the effect of the extra dimension is to average across it: for a wave in the fifth dimension the result of the averaging is zero. However the energy of the wave is positive because energy is the square of the wave. The (four-dimensional) energy-momentum tensor for the solution (62) is

$$T_{\nu\rho} = \frac{\omega c^3}{8\pi G_0} k_\nu k_\rho (\eta^{\sigma\tau} a_\sigma a_\tau) \cos^2(k_\mu x^\mu + k_5 x^5). \tag{63}$$

(Note that the corresponding expression in Ref. 1 is missing a factor of 2.) Since $T_{00}$ is positive the vector particle acts as a source of gravity with positive mass. Despite the resemblance to a regular photon, the solution (62) would seem to represent "cold" dark matter: $k_\mu$ is a timelike vector, and a particle described by this solution can have any sub-light speed – or even be at rest. Apart from the wave component, the energy-momentum tensor is the same as for ordinary matter, and therefore the equation of state for dark matter is likely to be the same as for ordinary matter. However it does not seem likely that there would be clustering of such particles, since these are waves, and waves interfere when they meet – there is no friction between them. Thus there are no dark matter planets or other such bodies in this model. Similarly, this dark matter neither emits nor absorbs light, because it is itself a wave.

The qualitative description of dark matter in this section also applies to the solutions given in subsection H and Sec. VI.

### G. Massive Vector-Tensor Particle

In addition to the solution (62) there is a particle which combines vector and tensor elements:

$$\varepsilon_\mu = 0$$

$$\varepsilon_5 = 1$$

$$\varepsilon^\mu = a^\mu \exp(ik_\rho x^\rho + ik_5 x^5)$$

$$g_{\mu\nu} = \eta_{\mu\nu} + a_{\mu\nu} \exp(ik_\rho x^\rho + ik_5 x^5),$$

(64)

where $g^{\mu\nu} k_\mu k_\nu + (k_5)^2 = 0$ and

$$g^{\mu\nu} k_\mu k_\nu a^\sigma + k_5 g^{\mu\nu} k_\mu a_{\nu\rho} g^{\rho\sigma} - (k_\mu a^\mu) k_\rho g^{\rho\sigma} - k_5 (a_{\mu\nu} g^{\mu\nu}) k_\rho g^{\rho\sigma} = 0,$$ (65)

$$2(k_\mu a^\mu) + k_5 a_{\mu\nu} g^{\mu\nu} = 0,$$ (66)

and

$$g^{\lambda\mu} g^{\nu\sigma} g^{\rho\tau} (k_\lambda k_\nu a_{\mu\rho} + k_\lambda k_\rho a_{\mu\nu} - k_\nu k_\rho a_{\lambda\mu})$$
$$- g^{\nu\sigma} k_5 k_\nu a^\tau - g^{\nu\tau} k_5 k_\nu a^\sigma = 0.$$ (67)

Equations (65), (66) and (67) can be solved with $a_{\mu\nu} = -(a_\mu k_\nu + a_\nu k_\mu)(k_5)^{-1}$ and $a^\mu = g^{\mu\nu} a_\nu$ giving solution

$$\varepsilon_\mu = 0$$

$$\varepsilon_5 = 1$$

$$\varepsilon^\mu = g^{\mu\nu} a_\nu \exp(ik_\rho x^\rho + ik_5 x^5)$$

$$g_{\mu\nu} = \eta_{\mu\nu} - (a_\mu k_\nu + a_\nu k_\mu)(k_5)^{-1} \exp(ik_\rho x^\rho + ik_5 x^5),$$

(68)

where $g^{\mu\nu} k_\mu k_\nu + (k_5)^2 = 0$. In the special case where $g^{\mu\nu} a_\mu k_\nu = 0$ the energy-momentum tensor is zero.

### H. Massive Graviton

Finally, there is also a solution where $\varepsilon^\mu = 0$, representing a massive graviton:

$$\varepsilon_\mu = 0$$

$$\varepsilon_5 = 1 \tag{69}$$

$$\varepsilon^\mu = 0$$

$$g_{\mu\nu} = \eta_{\mu\nu} + a_{\mu\nu} \exp(ik_\rho x^\rho + ik_5 x^5),$$

where $g^{\mu\nu}k_\mu a_{\nu\rho} = 0$, $a_{\mu\nu}g^{\mu\nu} = 0$ and $g^{\mu\nu}k_\mu k_\nu + (k_5)^2 = 0$. The corresponding energy-momentum tensor is

$$T_{\nu\rho} = \frac{c^3}{32\pi G_0} k_\nu k_\rho (\eta^{\sigma\lambda}\eta^{\tau\upsilon} a_{\sigma\tau} a_{\lambda\upsilon}) \cos^2(k_\mu x^\mu + k_5 x^5). \tag{70}$$

A non-zero graviton mass was first proposed by Fierz and Pauli[11], and a massive graviton appears in addition to a massless one in bimetric gravity[12]; see also the review articles[13,14,15]. Recently it has been considered whether a massive graviton could be a candidate for dark matter[16,17,18,19,20]. What distinguishes the theory of degenerate metrics from bimetric gravity is that in the degenerate theory the massless graviton and the massive graviton are different aspects of the same field, whereas in bimetric gravity they are separate fields.

Note that there is no massive scalar particle, for if $\varepsilon_5 = 1 + a\exp(ik_\rho x^\rho + ik_5 x^5)$ then the exponential term can be removed with an eigengauge transformation.

## VI. IMPOSING THE REFLECTION SYMMETRY

Under the reflection symmetry, considering weak fields, when going from charge described by $p_5$ to charge described by $u^5$, $\varepsilon_\mu$ is replaced by $g_{\mu\nu}\varepsilon^\nu$, changing the sign and dividing by $\sqrt{2\omega+1}$, while $\varepsilon^\mu$ is replaced by $g^{\mu\nu}\varepsilon_\nu$, changing the sign and multiplying by $\sqrt{2\omega+1}$; $\varepsilon_5$ and $g_{\mu\nu}$ are unchanged. When going from charge as $u^5$ to charge as $p_5$ the same procedure applies. Thus, for the four four-dimensional wave solutions, the scalar particle and the graviton are consistent with the symmetry because the vector fields are zero. The two photons are also consistent with the symmetry because they are independent of each other: if charge is described by $p_5$ then there is one photon which interacts with the particle (the covariant photon) and one which does not (the contravariant photon); if charge is described by $u^5$ then there is also one photon which interacts with the particle (the contravariant photon) and one which does not (the covariant photon). As for dark matter, the massive graviton is consistent with the reflection symmetry because both vector fields are zero. However neither the massive vector particle nor the massive vector-tensor particle is consistent with the symmetry. On its own the massive vector-tensor particle is not consistent with the reflection symmetry because one vector field is zero while the other is non-zero. In Ref. 1 it was stated that the massive vector particle was consistent with the symmetry, however the constants are not correct: there is a factor of $\sqrt{2\omega+1}$ missing. But it is possible to combine solutions (62) and (68) to obtain a new solution that is consistent with the reflection symmetry. This combined solution is

$$\varepsilon_\mu = -a_\mu \exp(ik_\rho x^\rho + ik_5 x^5)$$

$$\varepsilon_5 = 1 \tag{71}$$

$$\varepsilon^\mu = \sqrt{2\omega+1}\, g^{\mu\nu} a_\nu \exp(ik_\rho x^\rho + ik_5 x^5)$$

$$g_{\mu\nu} = \eta_{\mu\nu} - \left(\sqrt{2\omega+1}+1\right)(a_\mu k_\nu + a_\nu k_\mu)(k_5)^{-1} \exp(ik_\rho x^\rho + ik_5 x^5),$$

where $g^{\mu\nu}a_\mu k_\nu = 0$ and $g^{\mu\nu}k_\mu k_\nu + (k_5)^2 = 0$; the energy-momentum tensor is

$$T_{\nu\rho} = \frac{\omega c^3}{8\pi G_0} k_\nu k_\rho (\eta^{\sigma\tau} a_\sigma a_\tau) \cos^2(k_\mu x^\mu + k_5 x^5) \,, \tag{72}$$

i.e. the same as for the massive vector particle. The existence of this solution raises the possibility that the reflection symmetry can be *imposed* on the five-dimensional waves; it would then apply generally, without restriction. As was noted in Ref. 1 the field equations appear to be incomplete in that they do not contain, for example, any terms in $(\partial_5 \partial_5 \varepsilon^\mu)$: instead the term $(\partial_5 \partial_5 \varepsilon_\mu)$ appears twice, in equations (54) and (55). It is possible, then, that it is necessary to impose the reflection symmetry in order to complete the field equations. Note that the reflection symmetry has just been made to hold for this one particular solution: a procedure for ensuring that the symmetry holds in general would have to be developed.

## VII. CONCLUSION

In this paper a complete set of wave solutions has been found in the theory of degenerate metrics, assuming weak fields. It has been shown that the reflection symmetry can be imposed on dark matter by combining two of the solutions, with the implication that it may be possible to make the symmetry apply generally without compromising the new physics.

Imposing the reflection symmetry on the field creates a unusual technical problem, however, because it effectively introduces some new field equations, meaning that individual fields may be required to satisfy multiple equations. This is not a problem in itself: it has been demonstrated in this paper that solutions do exist to such a set of equations. However the analysis here has been restricted to weak fields, for which the field equations are linear. It remains to be seen whether the reflection symmetry can be applied exactly when one goes beyond weak fields to the full case with non-linear equations.